\documentclass[A4paper,11pt]{article}
\usepackage[english]{babel}
\usepackage{hyperref}
\usepackage[pdftex]{graphicx}

\usepackage[numbers]{natbib}
\usepackage{xcolor}
\usepackage{url} % Provides better formatting of URLs.
\usepackage{booktabs} % Allows the use of \toprule, \midrule and \bottomrule in tables for horizontal lines
\usepackage{graphicx}
\usepackage{siunitx}
\usepackage{xspace}
\newcommand{\GEANT}{{\textsc{Geant4}}\xspace}
\newcommand{\PadmeMC}{{\textsc{PadmeMC}}\xspace}
\newcommand {\ie}{\mbox{i.e.}\xspace}     %i.e.
\newcommand {\eg}{\mbox{e.g.}\xspace}     %e.g.
\newcommand {\vs}{\mbox{\sl vs.}\xspace}      %vs.

\begin{document}
\pagenumbering{arabic}
\titlepage {
\title{The PADME beam line Monte Carlo simulation}

\author{F.~Bossi$^{a}$,
P.~Branchini$^{b}$,
B.~Buonomo$^{a}$,
V.~Capirossi$^{c}$,
A.P.~Caricato$^{d,e}$, \\
G.~Chiodini$^{e}$,
R.~De Sangro$^{a}$,
C.~Di Giulio$^{a}$,
D.~Domenici$^{a}$,
F.~Ferrarotto$^{f}$, \\
S.~Fiore$^{g}$,
G.~Finocchiaro$^{a}$,
L.G~Foggetta$^{a}$,
A.~Frankenthal$^{h}$,
M.~Garattini$^{a}$,
G.~Georgiev$^{i,j}$, \\
A.~Ghigo$^{a}$,
P.~Gianotti$^{a}$,
F.~Iazzi$^{c}$,
S.~Ivanov$^{j}$,
Sv.~Ivanov$^{j}$, \\
V.~Kozhuharov$^{j,a}$,
E.~Leonardi$^{f}$,
E.~Long$^{k}$,
M.~Martino$^{d,e}$,
I.~Oceano$^{d,e}$, \\
F.~Oliva$^{d,e}$,
G.C. Organtini$^{k}$,
F.~Pinna$^{c}$,
G.~Piperno$^{k}$,\\
M.~Raggi$^{k,f}$\thanks{corresponding author: mauro.raggi@uniroma1.it},
%\note{Corresponding author}
I.~Sarra$^{a}$,
R.~Simeonov$^{j}$,
T.~Spadaro$^{a}$,
S.~Spagnolo$^{d,e}$,
E.~Spiriti$^{a}$,\\
D.~Tagnani$^{b}$,
C.~Taruggi$^{a}$,
P.~Valente$^{f}$,
A.~Variola$^{f}$,
E.~Vilucchi$^{f}$ 
\vspace{1cm}
\\ \\
$^a$INFN Laboratori Nazionali di Frascati, via E. Fermi 54, Frascati, Italy \\
$^b$INFN sez. Roma 3, via della vasca navale 84, Roma, Italy\\
$^c$DISAT Politecnico di Torino and INFN sez. Torino, C.so Duca degli Abruzzi 24, Torino, Italy\\ 
$^d$Dip. Mat. e Fisica Salento Univ., via Provinciale per Arnesano, Lecce, Italy\\ 
$^e$INFN sez. Lecce, via Provinciale per Arnesano, Lecce, Italy\\ 
$^f$INFN sez. Roma 1, p.le A. Moro 2, Rome, Italy\\ 
$^g$ENEA Frascati, via E. Fermi 45, Frascati, Italy\\ 
$^h$Physics Dept., Princeton Univ., Washington Road, Princeton, USA\\ 
$^i$INRNE Bulgarian Accademy of Science, 72 Tsarigradsko shosse Blvd., Sofia, Bulgaria\\ 
$^j$Sofia Univ. ``St. Kl. Ohridski'', 5 J. Bourchier Blvd., Sofia, Bulgaria\\ 
$^k$Dip. di Fisica Sapienza Univ., p.le A. Moro 2, Rome, Italy\\ 
}
% e-mail addresses: one for each author, in the same order as the authors
%\email{mauro.raggi@uniroma1.it}
%\emailAdd{second@asas.edu}
%\emailAdd{third@one.univ}
%\emailAdd{fourth@one.univ}
\maketitle

\abstract{
The PADME experiment at the DA$\Phi$NE Beam-Test Facility (BTF) of the INFN
Laboratory of Frascati is designed to search for invisible decays of dark 
sector particles produced in electron-positron annihilation events with a positron beam and a thin fixed target, by measuring the missing mass of single-photon final states. The presence of backgrounds originating from beam halo particles can significantly reduce the sensitivity of the experiment. To thoroughly understand the origin of the beam background contribution, a detailed \textsc{Geant4}-based Monte Carlo simulation has been developed, containing a full description of the detector together with the beam line and its optical elements. This simulation allows the full interactions of each particle to be described, both during beam line transport and during detection, a possibility which represents an innovative way to obtain reliable background predictions. }
%In this paper we 
%describe the details of this simulation 
%and report on the results obtained in 
%reducing beam background and 
%characterizing beam parameters. 
%detector and beam line description. 
%\end{Abstract}
}

%\begin{document} 
%\maketitle
%\flushbottom

\section{Introduction}

The PADME experiment~\cite{Raggi:2014zpa,Raggi:2015gza} at the DA$\Phi$NE Beam-Test Facility (BTF) of the INFN Laboratory of Frascati (LNF) is designed to detect invisible decays of dark sector particles produced in positron on fixed target annihilation, by measuring the missing mass of single-photon final states. The experiment is equipped with a \SI{100}{\um}-thick active diamond target~\cite{Oliva:2019alx} which was struck by a positron beam with energy of \SI{490}{\MeV} in Run I (November 2018 to February 2019) and \SI{430}{\MeV} in Run II (2020) data-taking periods. Non-interacting positrons are deflected by a dipole magnet, while photons produced in annihilation are detected by a calorimeter (ECal) made of 616~BGO crystals~\cite{Albicocco:2020vcy}. The crystals are arranged in a cylindrical shape, with a central square hole, necessary to avoid the overwhelming Bremsstrahlung photon rate at small angles.
%The hole acceptance is covered by a smaller, faster PbF$_{2}$-based calorimeter, specifically designed to reject the $e^+e^-\to \gamma 
%\gamma (\gamma)$ background.
%To reject the dominant Bremsstrahlung $\gamma$ background,
%a set of segmented plastic scintillator veto bars are used to detect irradiating positrons, 
%exiting the target with a lower momentum with respect to the original beam one.
To study detector performance, particle rates, signal acceptance and beam backgrounds, the full layout of the experiment is modeled using the \GEANT~\cite{GEANT4:2002zbu} simulation library. Run I used secondary positrons produced at the BTF target (upstream of the experimental hall) by high-energy electrons, followed by a few hours of running with primary positrons coming directly from the LINAC. 

The original experiment simulation did not include any description of the beam line upstream of the PADME target. Instead, nominal beam parameters were used and particle tracking began at our target only. During this first running period, a non-negligible beam-related background component was discovered in data. Therefore an additional part of the beam transport line was added to the \GEANT description of the experimental setup. The updated setup was used to study the origin of this beam-related background and thereafter to optimize the beam line configuration for Run II, at a slightly lower energy and using the primary positron beam only. In this paper, we describe in detail the recent implementation of the beam line description and the main results achieved using the simulation, while highlighting the innovative way in which we simulate both accelerator transport and experimental detection in a single simulation program.

%\section{The G4 implementation of the detector description}

% An example of a floating figure using the graphicx package.
% Note that \label must occur AFTER (or within) \caption.
% For figures, \caption should occur after the \includegraphics.
% Note that IEEEtran v1.7 and later has special internal code that
% is designed to preserve the operation of \label within \caption
% even when the captionsoff option is in effect. However, because
% of issues like this, it may be the safest practice to put all your
% \label just after \caption rather than within \caption{}.
%
% Reminder: the "draftcls" or "draftclsnofoot", not "draft", class
% option should be used if it is desired that the figures are to be
% displayed while in draft mode.
%
%\begin{figure}[!h]
%\centering
%\includegraphics[width=0.8\columnwidth]{trilece} 
%\caption{Simulation Results}
%\label{fig_sim}
%\end{figure}

% Note that IEEE typically puts floats only at the top, even when this
% results in a large percentage of a column being occupied by floats.

\section{The PADME Experiment Monte Carlo simulation} \label{sec:Exp-MC}

A \GEANT-based Monte Carlo (MC) simulation of the full experiment, called \PadmeMC, was developed in the early stages of the project to obtain first sensitivity estimates for the dark photon search~\cite{Raggi:2014zpa}. The \PadmeMC simulation package has evolved since then into a complete framework, capable of implementing complex data output structures, more detailed descriptions of the detector geometry, and different configurations of the setup. The software development closely followed the design and construction of the experiment~\cite{Leonardi:2017lxh} and was largely exploited to define the detailed design of the experiment and the various sub-detectors. In particular, it has been used extensively to verify the impact that different proposed technical solutions would have on the resolution of the dark photon recoil mass measurement and to optimize the construction parameters. 
%The current version of PadmeMC incorporates all of the 
%construction geometrical surveys and actual detector 
%geometry, together with most part of
%passive materials in the setup, such as supports, cables, shielding elements, etc.. 

During Run I, the control of beam-related backgrounds was understood to be of crucial importance for the sensitivity of the experiment. The total energy deposit in the PADME ECal due to beam halo photons needs to be below \SI{\sim 20}{keV/e^+} to avoid spoiling this sensitivity. Because this is an extremely challenging task for the BTF beam transport line when combined with the request of having \num{3e4} positrons in \SI{280}{\ns} long bunches, a detailed beam transportation simulation is essential. At the end of the run, \PadmeMC was improved to include a full simulation of the BTF beam transport line. The simulation now features the two bending dipoles, the two focussing/defocussing quadrupole pairs, and the beam collimators. 
%The quadrupoles, which are the main components of the transport line optics, have been simulated using the ??G4Quad class??.
%The code is currently available on GitHub.

%\subsection{Detector description}
%Describe how the geometry is implemented.

In the standard running conditions of the experiment, each LINAC pulse reaching the target is filled with \numrange[scientific-notation = fixed]{25e3}{30e3} positrons, with mm-scale beam spot size and a few per-mil beam energy resolution. The beam intensity is thus defined by the number of positrons on target (NPOT). The beam bunches have length ranging from \SI{250}{ns} to  roughly \SI{300}{ns}, depending on the configuration of the LINAC gun and the accelerating RF in its four power stations, with a fine micro-bunch structure given by the \SI{2856}{MHz} of the RF. The repetition rate is \SI{50}{Hz}: in one second 49 pulses are delivered to the BTF line, and one is diverted by a pulsed dipole to a spectrometer line for monitoring the central value and spread of the beam momentum. 

To simulate the interaction of beam particles with the active target, events with user-defined NPOT can be generated and transported into the BTF beam line.  
%The event kinematics resulting from the interaction of these particles with the target are simulated using the standard GEANT4 ??EM?? physics libraries. 
%A realistic description of the incoming positron beam is crucial for a 
%reliable evaluation of the resolution in the measurement of the $A'$ recoil 
%mass and exotic signal acceptance. For this reason, the PADME 
%MonteCarlo has been tuned on data coming from dedicated 
%runs collected during PADME RunI and RunII.
To cope with the different running conditions of PADME Run I and Run II, the simulation allows for the tuning of all relevant beam parameters:
\begin{itemize}
    \item Total duration and internal time structure of the positron bunch;
    \item Energy spread, spatial distribution, beam emittance, and energy spread of the beam spot at the target.
\end{itemize}

%The beam simulation package also includes methods to produce special events for the calibration of
%the electromagnetic calorimeter where fixed energy photons are directed to specific areas of the
%detector

\subsection{Simulation physics}

The PADME list of simulated physical processes is derived from the standard QGSP\_BERT physics list provided by the \GEANT package. It includes multiple scattering, Coulomb scattering (Bhabha S and T channels), ionization, Bremsstrahlung emission, two-photon annihilation, synchrotron
radiation emission, and, optionally, tracking of optical photons. Specific datacards allow the inclusion of photonuclear interactions and the selection of the high-precision neutron transport library to use.
%\color{blue} Venelin please check this
%\color{black}

Simulation of exotic particle production, such as annihilation with dark photon emission $e^+e^- \to A'\gamma$, is not part of the \GEANT physics package, and is handled instead by custom generators configurable via datacards. Higher-order radiative corrections to electromagnetic processes are also not implemented in \GEANT, but can produce relevant background in searches for rare processes such as dark particle production. In the PADME Monte Carlo simulation, the kinematics of the three-gamma final state $e^+e^- \to \gamma\gamma(\gamma)$ is produced externally to \GEANT using the \textsc{CalcHEP}~\cite{Belyaev:2012qa} and \textsc{BabaYaga}~\cite{Balossini:2008xr,Balossini:2006wc} generators.

%\subsection{Output description}

\section{The PADME beam line Monte Carlo simulation} \label{sec:beam-line}

The problem of conciliating the simulation of beam dynamics with beam-related backgrounds is of crucial importance to the larger HEP community, in particular for future very high-energy and intense beams. Few solutions to this problem exist; one example is \textsc{G4BeamLine}~\cite{Roberts:2008zzc}, which is not commonly used by the particle physics community. Recently, the \textsc{BDSim} framework, also based on \GEANT, has been gaining momentum~\cite{Nevay:2018zhp}. Although some beam line simulations using \GEANT were successfully realized in the past (\eg, the HARP experiment),
%, in the PADME beam line Monte Carlo the full description of the transport line and of the experimental setup is performed in within the same GEANT4 program for the first time. 
in PADME we have achieved for the first time a simultaneous simulation within a single \GEANT program of both beam line and experiment, containing a full description of the transport line.

%Data of PADME Run I clearly demonstrated that the beam simulation starting from the experiment target, was unable to predict the
%observed background in the PADME detectors

\begin{figure}[ht]
\centering
\includegraphics[width=0.85\linewidth]{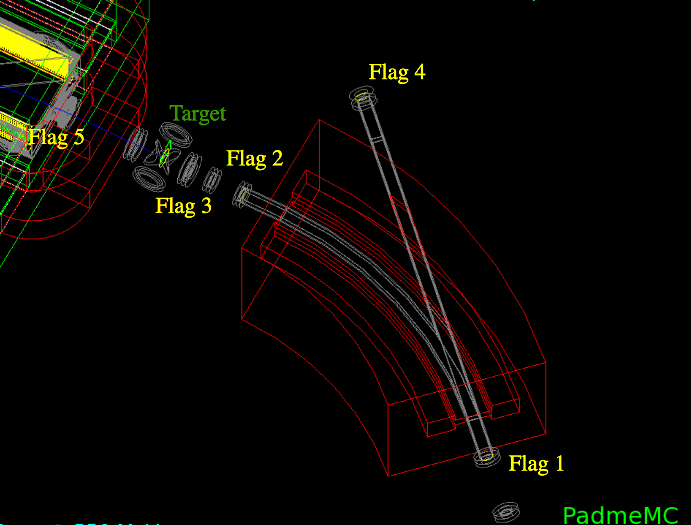}
\caption{Detailed view of the PADME target region: in green Target with its support, in yellow the beam ``Flags'', and in red the BTF DHSTB002 dipole magnet.}
\label{fig:Flags}
\end{figure}

When compared to the PADME experimental data of Run I, simulations starting only from the target interaction were unable to reproduce the observed background, which was dominantly produced from the beam halo interaction with beam line materials. To identify the sources of such a beam background, a complete description of the last \SI{\sim 15}{m} of the transfer line of the BTF was implemented in the simulation, adapting existing \GEANT classes.
%For the first time the optics of the line has been implemented using G4Quad, for the quadrupoles 
%description and a magnetic field map for the bending dipoles.
On the beam line, a set of virtual detectors called ``Flags'' (see Fig.~\ref{fig:Flags}) was introduced to monitor the effect of the optical elements on the beam spot size and shape.

\subsection{Run I beam line simulation}

\begin{figure}[ht]
\centering
 \includegraphics[width=1.\linewidth]{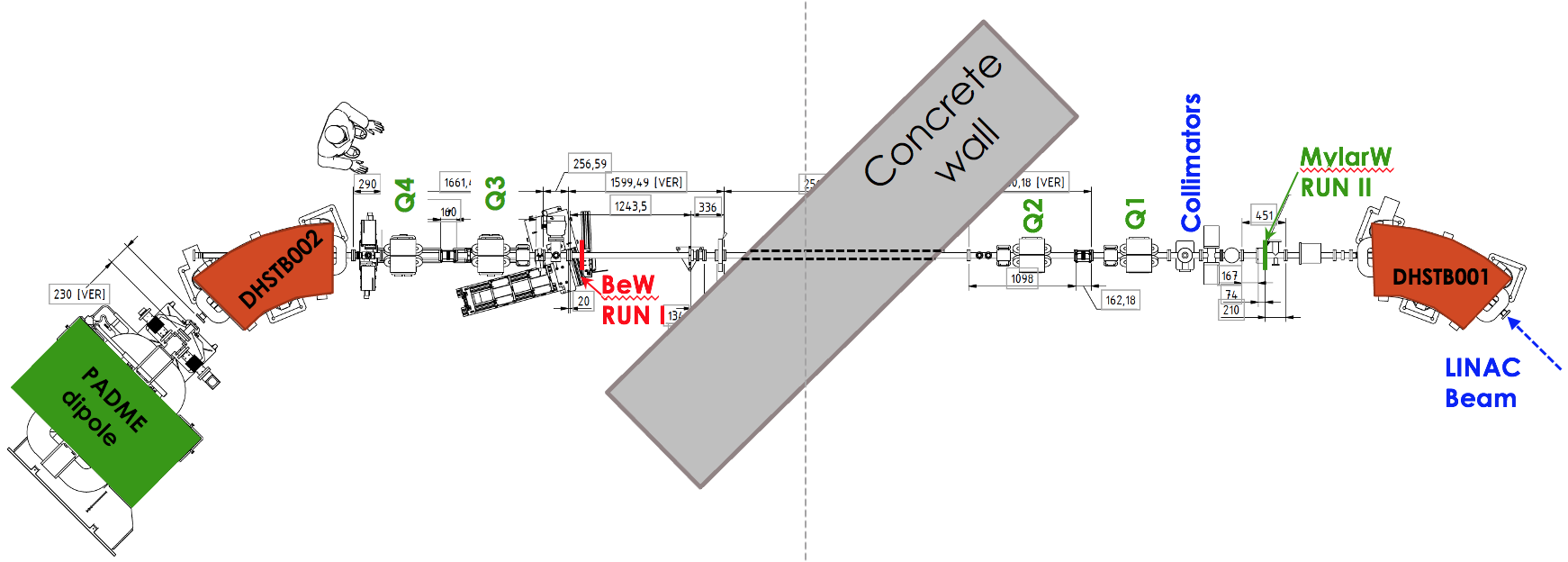}
\caption{Drawing of the PADME beam line setup during Run II in 2020.}
\label{fig:Outline_2020Line}
\end{figure}

During Run I the beam transport line entering the PADME experiment was separated from the LINAC vacuum by means of a \SI{250}{\micro\meter} thick Berillium (Be) window, to protect the LINAC from possible vacuum leaks in the experimental setup. The window was placed inside the BTF hall just before the last quadrupole pair. 
%bending magnet DHSTB002. 
This location is represented by a red line in Fig.~\ref{fig:Outline_2020Line}. The Be window was suspected as the origin of the main component of the observed beam-related background, and therefore the line was initially simulated starting from this position. 
%as shown in Fig. \ref{fig:Flags}.

\subsection{Run II beam line simulation}

%Based on the result of the Run I backgrounds simulations, the PADME Run II beam line was modified to reduce its contribution to the experiment background. 
To reduce the background on the detector, the PADME beam line was modified for Run II based on the simulation results. The main interventions were:

\begin{itemize}
    \item The Be window was removed and replaced with a \SI{125}{\micro\meter}-thick MYLAR window positioned \SI{\sim 10}{m} upstream at the exit of the DHSTB001 magnet (green line in Fig.~\ref{fig:Outline_2020Line});
    \item The clearance of all beam pipes in the line was increased to \SI{60}{mm} by removing a pulsed magnet at the entrance of the BTF hall~\cite{Foggetta:2021gdg}.
\end{itemize}

The main elements of the updated line were implemented in the new PADME beam line simulation, including the optics of the two pairs of quadrupoles ($Q_1$ $Q_2$ and $Q_3$ $Q_4$).

\begin{figure}[ht]
\centering
\includegraphics[width=\linewidth]{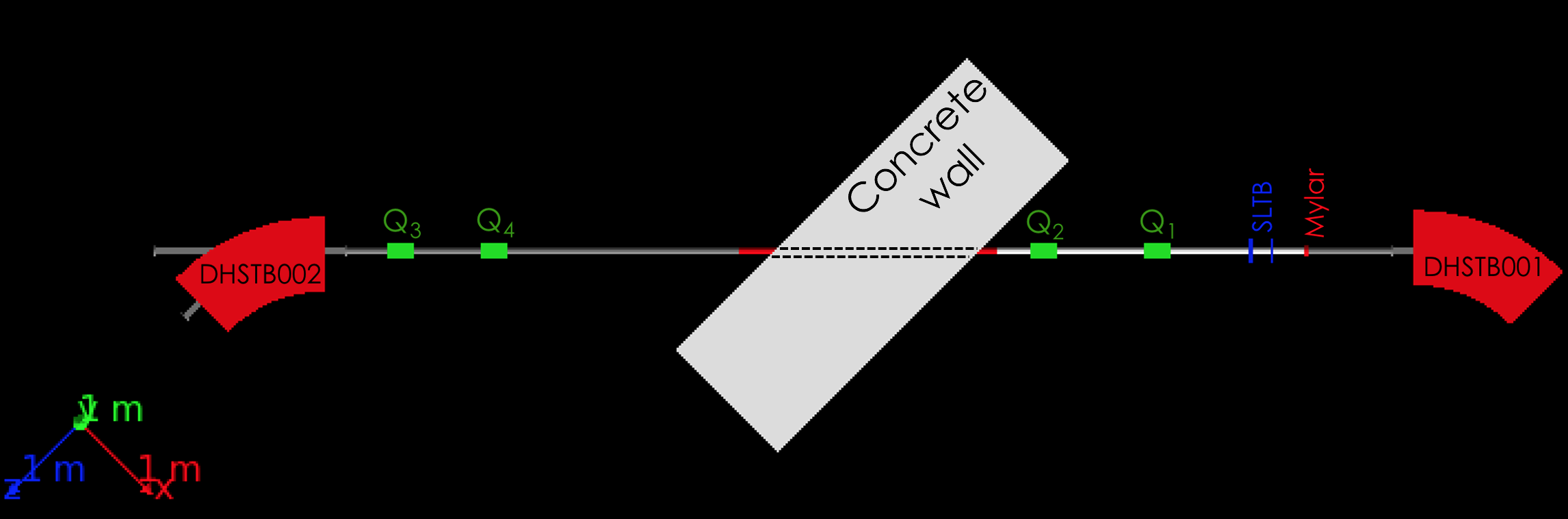}
\caption{Top view of the Monte Carlo description of the PADME beam line setup during Run II.}
\label{fig:2020Line}
\end{figure}

In Fig.~\ref{fig:2020Line} the dipole magnets DHSTB001 and DHSTB002 are shown in red, the four quadrupoles (Q1-Q4) in green, the collimators (SLTB) in blue, and the MYLAR window flange also in red. To get a better understanding of the beam halo background, the two collimators SLTB3 (vertical, placed \SI{309}{mm} from the MYLAR window) and SLTB4 (horizontal, placed \SI{519}{mm} from the MYLAR window) were also simulated to more accurately reproduce the actual beam shape and energy resolution. The concrete wall separating the LINAC tunnel from the BTF hall is also simulated, to screen the experiment from background generated by the interaction between beam halo and collimators.

\section{Beam line Monte Carlo simulation results}

The \PadmeMC beam line simulation has been used to study several aspects of the experiment: 
\begin{itemize}
%    \item the electromagnetic calorimeter performance comparison between real data from RunI and RunII and simulation results using $e^+e^- \to \gamma\gamma$ as benchmark process.
    \item Beam background levels during Run I and Run II;
    \item Effect of the MYLAR window thickness;
    \item Quadrupole settings of the transport line;
    \item Beam energy resolution;
    \item Absolute value of the positron beam energy.
\end{itemize}

%\subsection{ECal performance}

%The BGO based electromagnetic calorimeter (ECal) is the most important 
%detector of the PADME experiment, and the assessment and optimization of its
%performance are of the utmost importance. The best way to study the ECal 
%performance is to select the benchmark process $e^+e^- \to \gamma\gamma$, and the reconstruction the total Energy $E_{tot}=E_{\gamma1}+E_{\gamma2}$, which is equal to the beam energy. The comparison of the total energy resolution and tail distribution of the $E_{tot}$ spectrum will validate the MC capability of mimic the calorimeter performance together with the beam halo component related pile-up.

%\subsection{Run I and Run II beam background studies}

In the next section we describe the capability of the simulation to reproduce the background generated in the experiment by interactions of the beam halo with beam line components and passive materials in the experiment. We also show how the simulation helps to understand the origin of the beam halo and how a significant reduction has been achieved.

\subsection{Beam halo generation}

During the last days of Run I and for the entirety of Run II, the experiment used a primary positron beam, \ie, the beam came directly from the LINAC instead of being generated via the secondary target. This choice was aimed at minimizing the production of background photons in the experimental area. For the same reason, the use of collimators to select a narrow momentum band of \num{\sim 0.4}{\%} was limited to the region downstream of the first bending magnet DHSTB001 (see Fig.~\ref{fig:Outline_2020Line}), separated from the experimental area by a concrete wall more than a meter thick. All these conditions excluded the possibility that the photon background observed in the PADME calorimeter came from outside the beam pipe. 

\begin{figure}[ht]
\centering
\includegraphics[width=0.7\linewidth]{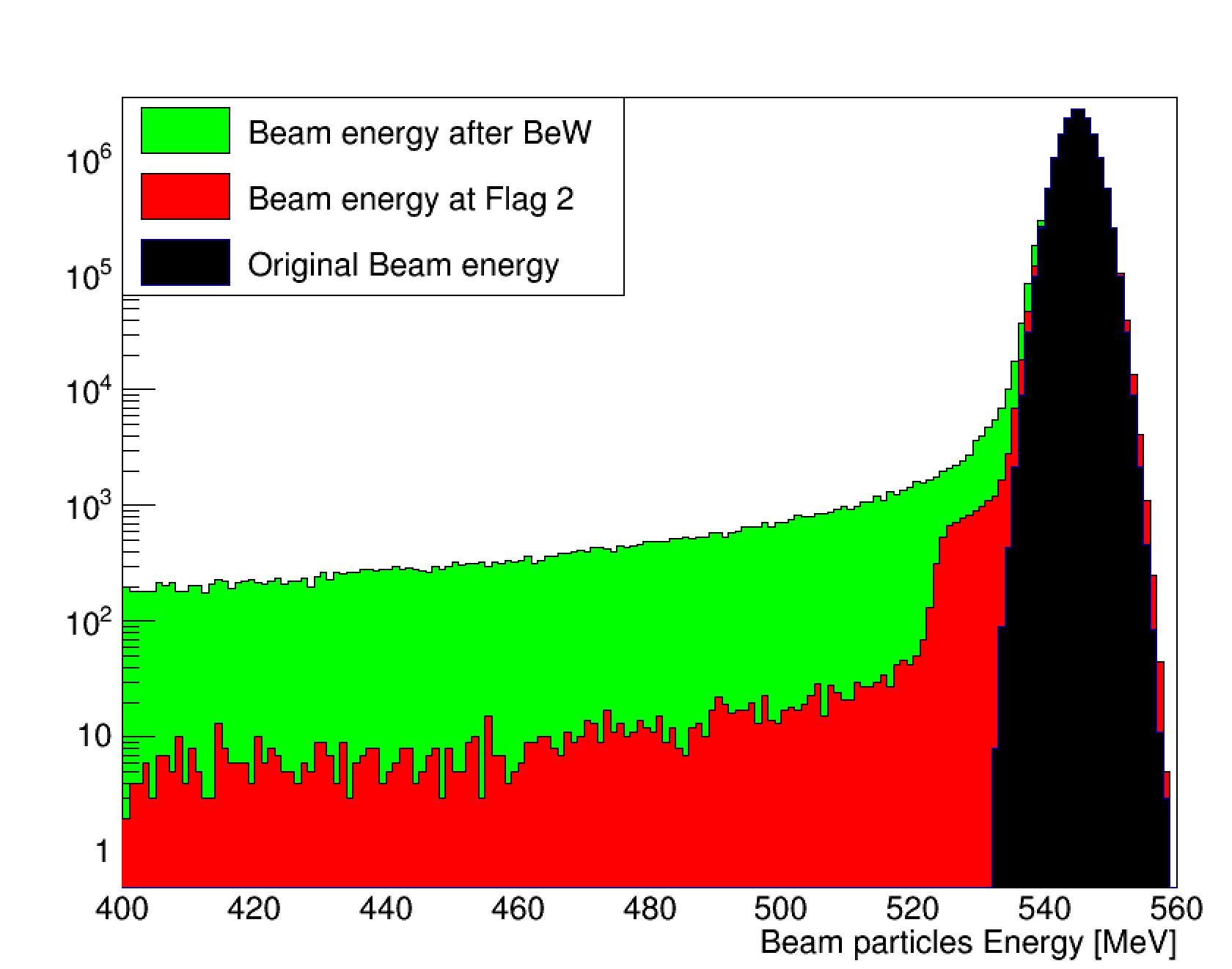}
\caption{Beam energy profile as seen along different sections of the beam line.}
\label{fig:EBeamPath}
\end{figure}

Using the Monte Carlo simulation of the beam transport along the line, the energy distribution of the beam was compared before and after the Be window. In a second step, the beam was also observed at the exit of the DHSTB002 dipole (Flag 2 in Fig.~\ref{fig:Flags}) just before the PADME target. Fig.~\ref{fig:EBeamPath} shows the effect of beam transport on the beam energy profile under these different conditions. Starting from the originally simulated pure Gaussian energy resolution of \num{0.5}{\%} (black distribution), the beam develops a low-energy tail after crossing the Be window (green distribution), which is partially cut off by the output flange of the DHSTB002 magnet (red distribution). The Be window thus generated significant Bremsstrahlung tails in the beam energy distribution, and the simulation also demonstrated that the low-energy tails were cut off due to particles hitting the border of the DHSTB002 exit vacuum pipe and connection flange.

\begin{figure}[ht]
\centering
\includegraphics[width=0.8\linewidth]{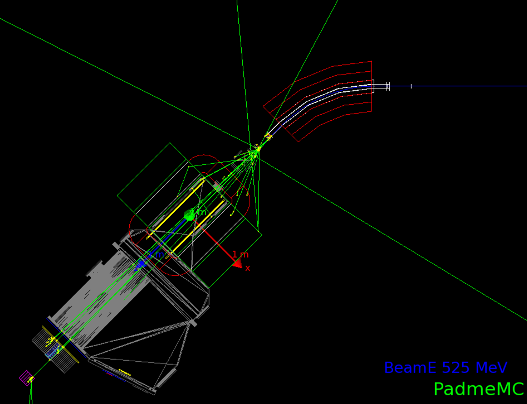}
%\caption{Top view of the PADME beam line MC for the RunI in early 2019}
\caption{Low-energy positron striking the DHSTB002 dipole exit flange, thereby generating a shower. Green lines represent photons.}
\label{fig:2019Linelarge}
\end{figure}

The low-energy tail corresponds to positrons losing a substantial fraction of their energy (\SI{> 25}{MeV}) in the interaction with the Be window, leaving the nominal beam trajectory and finally crashing onto the DHSTB002 exit vacuum pipe, as shown in Fig.~\ref{fig:2019Linelarge}. Some high-energy photons, generated by the interactions just described, were also able to reach the ECal by traveling inside the PADME vacuum system.

Therefore, before the start of Run II PADME decided to move the vacuum separation window farther upstream, closer to the DHSTB001 magnet, and to replace the \SI{250}{\micro\meter} Be window with a thinner MYLAR window.

\clearpage

\subsection{Effect of the MYLAR window thickness on beam backgrounds}

\begin{figure}[ht]
\centering
\includegraphics[width=0.7\linewidth]{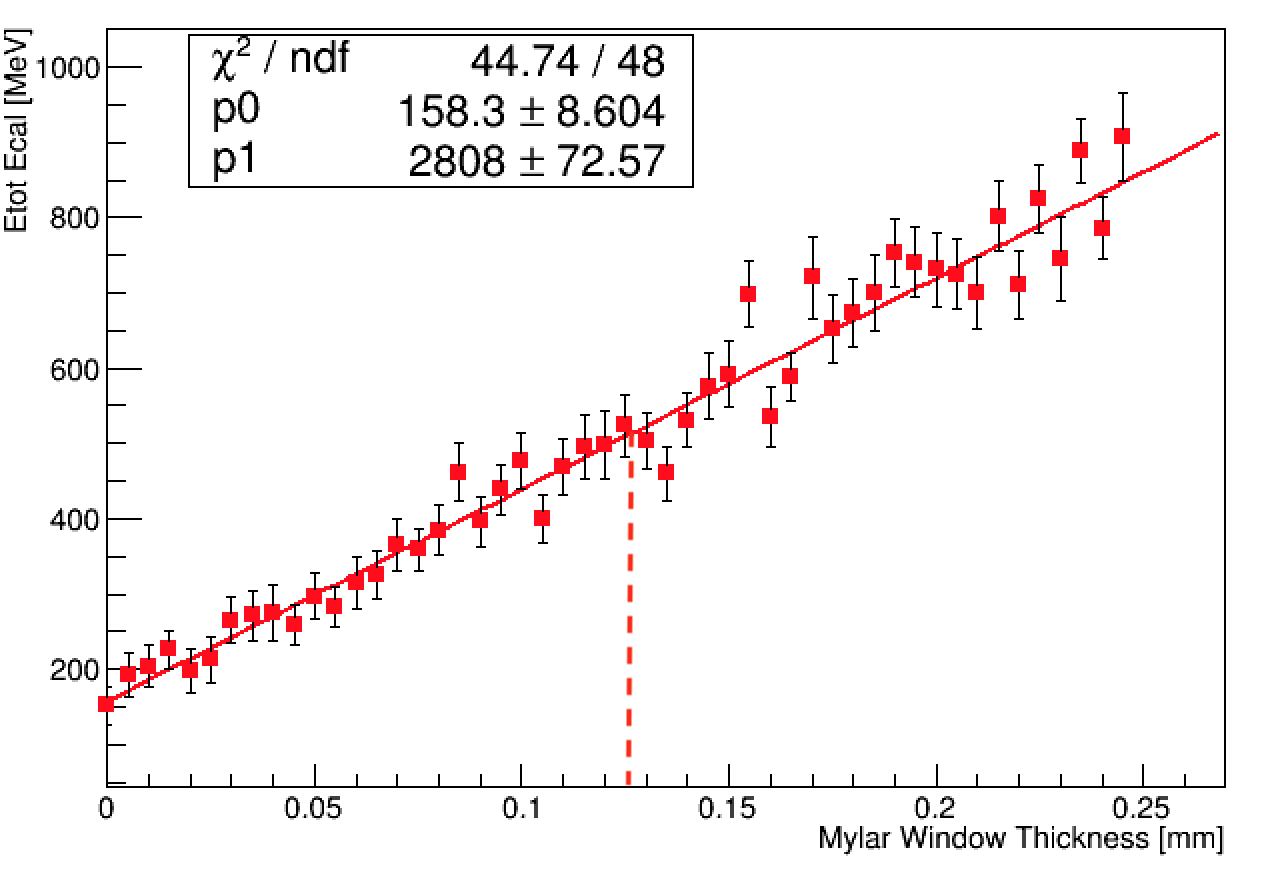}
\caption{Effect of the window thickness on the PADME experiment background for \num{25e3} positrons on target using the 2020 beam line configuration.}
\label{fig:WinThick}
\end{figure}

In Run II, the LINAC vaccuum was separated from the experiment vaccuum using a MYLAR window (shown as a green bar in Fig.~\ref{fig:Outline_2020Line}). The effect of the thickness of the window on the background energy observed in the PADME ECal was studied in order to find the maximum allowable thickness, which corresponds to the maximum allowable background. In Fig.~\ref{fig:WinThick} the total energy deposit in the calorimeter is shown as a function of window thickness in mm. The plot was obtained by simulating one thousand events of \num{25e3} POT each for every \SI{5}{\micro\meter} step in window thickness. The \SI{\sim 160}{\MeV} energy deposit observed for \SI{0}{\micro\meter} thickness is produced by the interaction of the beam with the diamond target. Using the simulation, and after laboratory tests of the window mechanical strength, the minimum window thickness was fixed to \SI{125}{\micro\meter} (red-dashed line). The simulation predicted an average energy deposit of \SI{\sim 500}{\MeV} per \num{25e3} POT, in very good agreement with the measurement performed during Run II data taking. 
 
\subsection{Beam background reduction}

\begin{figure}[ht]
\centering
%\includegraphics[width=1.\linewidth]{Figures/EtotDifferentRuns.png}
%\caption{Total deposited energy in PADME ECal during different running periods.}
\includegraphics[width=0.65\linewidth]{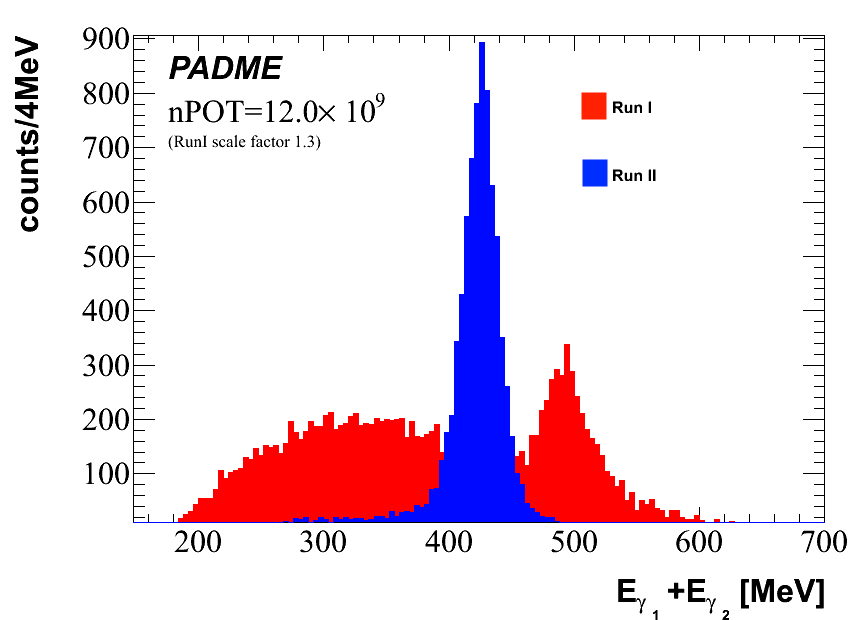}
\caption{Total energy of cluster pairs during different running periods.}
\label{fig:ECalETot}
\end{figure}

With the new beam line configured for Run II by the Beam Test Facility staff, a considerable reduction of the beam-related background was achieved. Fig.~\ref{fig:ECalETot} shows the total energy of two clusters for late Run I (red) and Run II (green) with different beam line configurations. Events were selected by requiring the two clusters to be within \SI{10}{ns} of each other; inside the fiducial region of the calorimeter; and with an energy center-of-gravity below \SI{5}{cm}. The peak in the distribution is produced by $e^+e^- \to \gamma\gamma$ and corresponds to the different beam energies: \SI{490}{\MeV} for Run I and \SI{430}{\MeV} for Run II. The low-energy region is dominated by pile-up background. The green distribution shows strong event suppression in the background region. This represents the main achievement obtained by moving the vacuum separation window and replacing \SI{250}{\micro\meter} of Be with \SI{125}{\micro\meter} of MYLAR. The improvement in the $\gamma\gamma$ peak resolution and the absence of high-energy tails in the Run II distribution (green) are additional consequences of the lower pile-up.  

\subsection{Data \vs Monte Carlo simulation comparison}

\begin{figure}[ht]
\centering
\includegraphics[width=0.65\linewidth]{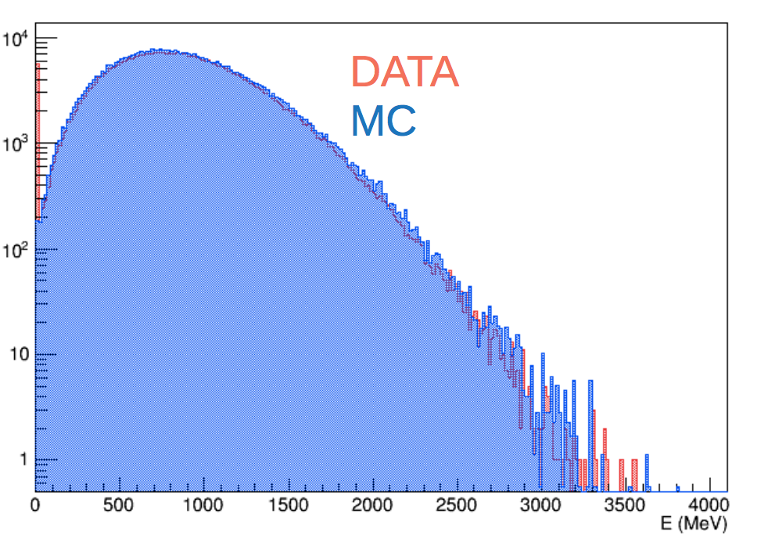}
\caption{Total cluster energy distribution during Run II, measured experimentally (red), and with the MC simulation (blue).}
\label{fig:DataMC}
\end{figure}

To test the quality of the background description achieved by the beam line MC simulation, we compared the distribution of the total cluster energy obtained from simulation with that obtained from data in Run II. The data sample was collected with standard beam conditions: \num{30e3} positrons on target and \SI{280}{ns} bunch length. The data distribution, in red in Fig.~\ref{fig:DataMC}, is well described by the simulated one in blue. Even in the high-energy region, where the MC slightly overestimates the data, the agreement is still good.

\subsection{Quadrupole gradient tuning}

To verify the quality of the optics simulated by the PADME beam line MC simulation, we performed a scan of quadrupole gradient fields. The BTF line uses two quadrupole pairs: Q1 and Q2 in the LINAC region, and Q3 and Q4 in the BTF hall (see Fig.~\ref{fig:2020Line}). The beam dimensions measured at the PADME target with simulation were compared to the ones measured experimentally. During Run II, the beam dimensions at the target were \SI{\sim 1.5}{mm} vertically and \SI{\sim 1.2}{mm} horizontally. After a first gradient scan, it was established that the shape of the beam at the target can be adjusted by changing only the gradient of the last two quadrupoles. Fixing the gradients of the first two quadrupoles Q1 and Q2 to the actual values used during Run II, we scanned the gradients of Q3 and Q4, searching for the values providing horizontal and vertical spot sizes of \SI{\sim 1.5}{mm}. The scan ranged from \SIrange{1}{5.2}{T/m} in 15 steps of \SI{0.3}{T/m}.

\begin{figure}[ht]
\centering
\includegraphics[width=1.\linewidth]{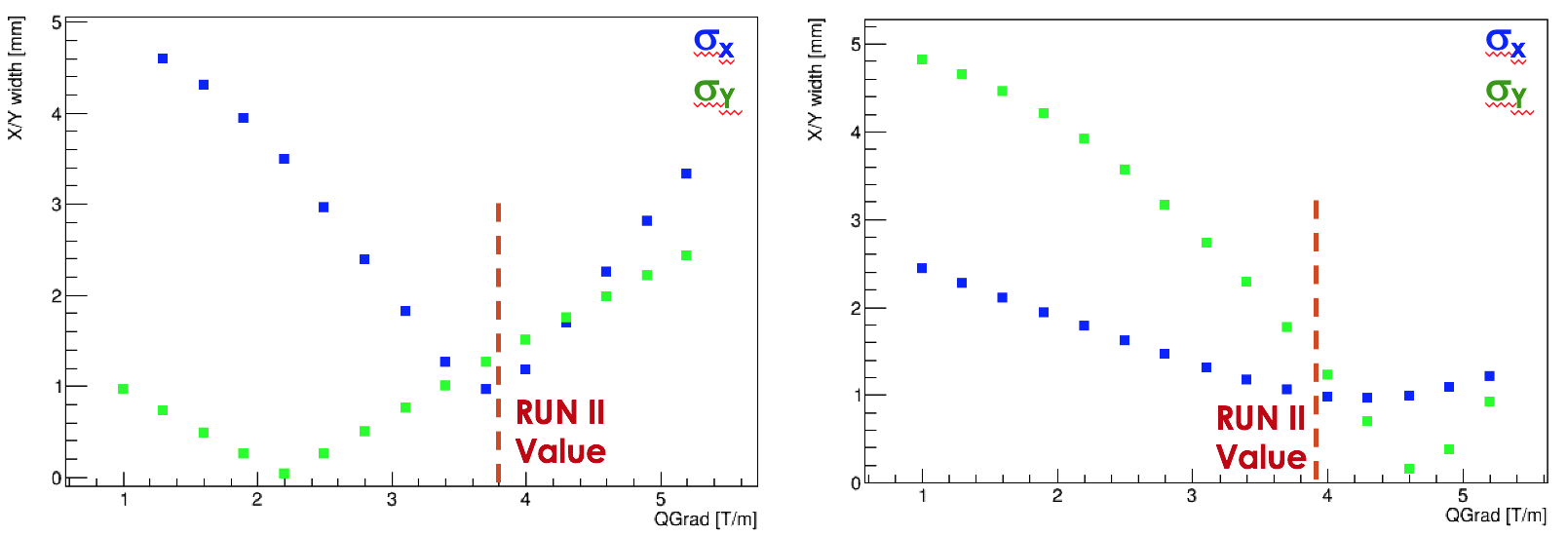}
\caption{Beam transverse width on target \vs quadrupole Q3 (left) and Q4 (right). The vertical red-dashed line represents the actual gradient used during Run II.}
\label{fig:QuadTune}
\end{figure}

In Fig.~\ref{fig:QuadTune}, the results of the scan are shown for Q3 and Q4. On the Y axis of the plot, the RMS of the beam spot size at the PADME target is represented for different values of the quadrupole gradient field. According to the scan, the best value for the Q3 gradient is roughly \SI{3.7\pm0.3}{T/m}, compared to the \SI{3.8}{T/m} actually used. For Q4, the set of possible gradients is somewhat larger but the value that leads to both X and Y RMS closest to \SI{1}{mm} is \SI{4 \pm 0.2}{T/m}, very close to the \SI{3.9}{T/m} actually used.

\subsection{Beam energy dispersion estimates}

One way to estimate the accepted beam energy range generated by the BTF transport line is to use the dispersion created by DHSTB001 and the aperture of the horizontal downstream collimator SLTB4. Using the Monte Carlo simulation of the beam bending through the DHSTB001 dipole, we derived the correlation coefficient $a$ of the impact position at the collimator \vs particle energy. 
%due to the dipole chromatic effect. 
Given the aperture of the horizontal collimator SLBT4, placed just after the DHSTB001 dipole, the energy dispersion of the PADME beam was estimated with the formula:

\begin{equation}
\left|\frac{\Delta E}{E}\right| = \Delta X \cdot a
\label{Eq:AbsEnOriginal}
\end{equation}

\noindent where $\Delta X$ is the horizontal collimator aperture in mm and $a =  \SI{0.267}{MeV/mm}$. Using the $\Delta X$ value set during Run II (\SI{\pm 2}{mm}), we obtain a nominal beam energy spread of \SI{\sim 1.1}{MeV}. Dividing by the beam energy of \SI{430}{MeV} then gives $\left|\frac{\Delta E}{E}\right|\sim 0.25\%$. The result is confirmed in simulation by studying the total amount of energy deposited in the ECal. After tuning the beam optics, we performed a simulation scan to measure the total energy in the calorimeter as function of beam energy spread, from \numrange{0}{0.75}{\%} in 15 steps. The background level in the ECal remains stable and compatible with that observed in Run II data as long as the beam energy spread stays below \num{0.4}{\%}. 

An independent estimate can be obtained analytically using Eq.~(3) of Ref.~\cite{Ghigo:2003gy}, which also takes into account the entrance angle at DHSTB001:

\begin{equation}
\left|\frac{\Delta E}{E}\right|=\frac{h}{2\rho}+\sqrt{2}\left(\frac{R_x}{L_1}+\frac{H}{2L_1}\right)\simeq\frac{h}{2\rho}+ \sqrt{2}\frac{H}{L_1}.
\end{equation}

Here, $\rho$ is the radius of the dipole, $L$ and $h$ are the distance and aperture of the downstream collimator SLTB4, and $H$ is the aperture of the upstream collimator SLTB2. In the BTF case, $\rho = \SI{1.723}{m}$ and $L_1 = \SI{1.4750}{m}$. Using the collimator apertures $h = \SI{4}{mm}$ and $H = \SI{1.7}{mm}$, typical values used during Run II, and a beam spot size $R_x = \SI{1}{mm}$, we obtain an estimated energy spread of $\left|\frac{\Delta E}{E}\right|\sim 0.3\%$, in very good agreement with the simulation estimate.

\subsection{Absolute measurement of the beam energy}

The description of the magnetic field of the PADME dipole in the experiment simulation is extremely accurate and uses a detailed field map obtained using remote-controlled Hall probes.
%by the LNF accelerator division. 
The PADME magnet is a CERN MBP-S dipole with an increased vertical gap of \SI{230}{mm}. 

To evaluate the impact on the magnetic field, the dipole magnetic volume was scanned at LNF in 3D in steps of a few mm in each of the three coordinates, including the fringe field regions. The excitation curve was also measured allowing the conversion of magnet current (I) into magnetic field (B) at the center of the coils: 

\begin{equation}
    \mathrm{B} = 19.44 \, \mathrm{\times \, I(A)} + 32.8 \quad \mathrm{[G]}.
\end{equation}

\noindent The dipole current was monitored by the PADME Detector Control System
(DCS) every few seconds during all of Run I and Run II.

\begin{figure}[ht]
\centering
\includegraphics[width=1\linewidth]{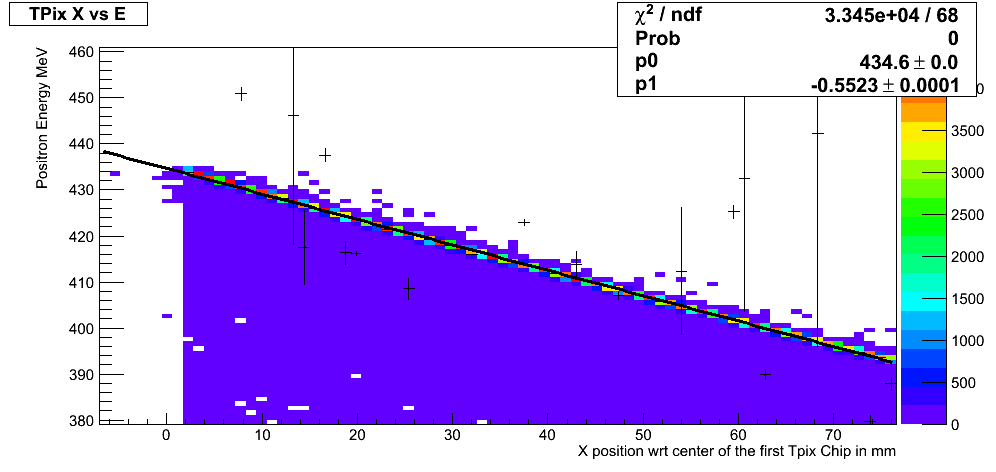}
\caption{Calibration of the positron beam impact point on the TimePix detector with the MC simulation.}
\label{fig:TPixCalib}
\end{figure}

With the beam position measured by the target before the dipole entrance and close to the beam dump by means of a TimePix detector \cite{Poikela_2014}, an absolute measurement of the beam energy can be obtained. The measured dipole magnet current and the beam impact position on the active target were fed to the simulation as input. To predict the impact point at the TimePix, simulation samples with different positron beam energies in the \SIrange{380}{435}{MeV} range were generated. Fig.~\ref{fig:TPixCalib} shows the correlation of the predicted impact point on the TimePix detector as a function of the primary positron energy. 

Profiling and fitting the distribution of Fig.~\ref{fig:TPixCalib} we obtain a function converting the beam impact point position on the TimePix detector (X) into a beam energy as follows:

\begin{equation}
E^{\text{MC}}(X \, \mathrm{[mm]}) = (434.64-0.5523 \times X \, \mathrm{[mm]}) \quad \mathrm{[MeV]}
\label{Eq:AbsEn}
\end{equation}

During a dedicated test, we measured the impact point with the TimePix detector to be \SI{-3.9}{mm} in local detector coordinates, with negligible error. Using Eq.~\ref{Eq:AbsEn} yields $E_{\text{meas}}^{\text{MC}} = \SI{432.9}{MeV}$, less than \SI{1.0}{MeV} away from the \SI{432.5\pm2.2}{MeV} measured with the LINAC hodoscope in the July 2020 technical run~\cite{Foggetta:2021gdg}. The measurement was obtained with the same LINAC settings and transfer line used during Run II. 
%Using horizontal
%scrapers, thinning the gap between them, we perform %the selection of the energy bin which lasts longer in the overall
%pulse envelope, via
The statistical errors due to the fit are small and the error on the position of the beam at the target is just \SI{\sim 100}{\micro\meter}~\cite{Oliva:2019alx}, while the position of the target itself is known to \SI{\sim 250}{\micro\meter} accuracy. The dominant systematic uncertainty comes from possible variations in the real position of the TimePix detector with respect to the simulation model, estimated to be \SI{\sim 2}{mm}. Another source of systematic error is the value of the fit parameters extracted in Fig.~\ref{fig:TPixCalib} and used in Eq.~\ref{Eq:AbsEn}. Changing the fit procedure and the fit binning, relative variations of \num{\sim 0.1}{\%} were observed. The final result for the PADME beam energy measurement in Run II is:\\

\begin{equation}
E_{\text{meas}}^{\text{MC}}=(432.9 \pm 0.1_{\text{stat.}}\pm{1.1}_{\text{syst.}}) \quad \mathrm{[MeV]}
\label{Eq:AbsEnMC}
\end{equation}
%The results is satisfactory and demonstrates that the simulation of the PADME dipole magnetic field is accurate enough to reproduce the impact position of charged particles, which is crucial to validate the use of the PADME vetos as charged particle spectrometers.

A separate method to determine the beam energy exploits the DHSTB002 dipole magnet bending and the position of the beam at the target. This method is currently less precise due to the unknown position of the beam at the entrance of DHSTB002, and the shorter distance between the DHSTB002 entrance and the target. In 2020, a new horizontal collimator was added at the entrance of DHSTB002, allowing us to fix the beam position with higher precision.

The PADME experiment plans a dedicated run to search for the recently postulated X17 particle \cite{Krasznahorkay:2015iga}. In fact, it has been pointed out in Ref.~\cite{Nardi:2018cxi} that using resonant production, $e^+e^- \to X17 \to e^+e^-$, it would be possible to improve the experiment sensitivity to X17. In this type of search, establishing the exact resonance mass ($\simeq \sqrt{m_eE_{beam}}$) is vital. Thus the ability to measure the beam energy with sub-percent precision is key. With the present energy determination method, and $\sqrt{s} = \SI{17}{MeV}$, corresponding to a beam energy of \SI{282}{MeV}, the uncertainty on the beam energy will be \num{0.25}{\%}, or \SI{\approx 1}{MeV}, which translates to a mass uncertainty of \SI{\approx 30}{KeV}.

Improvements that will help in further reducing the error on the energy measurement are still possible by, \eg, increasing the precision of the TimePix detector positioning and of the MC simulation magnetic field maps.

\section{Conclusions}

The PADME beam line Monte Carlo simulation implements a full \GEANT-based simulation of the DA$\Phi$NE Beam Test Facility transport line, from the end of the LINAC to the target of the PADME experiment. The simulation is able to reproduce the beam optics, to predict the correct quadrupole magnetic setting, and to account for the beam halo background observed in the experiment calorimeter. Furthermore, it has been used to measure the absolute beam energy ($432.9\pm 0.1 \pm 1.0$) MeV and its spread, \num{0.25}{\%}, with high accuracy. These improvements will allow the simulation to produce reliable background predictions in different configurations, helping to plan the PADME Run III setup.

\section{Acknowledgments}
We warmly thank all of the BTF and LINAC teams of Laboratori Nazionali di Frascati for providing an excellent quality beam and full support during data taking and simulation development. 

This work is partly supported by the Italian Ministry of Foreign Affairs and International Cooperation (MAECI) under the grant PRG00226 (CUP I86D16000060005), the BG-NSF KP-06-DO02/4 from 15.12.2020 as part of MUCCA, CHIST-ERA-19-XAI-009, and TA-LNF as part of STRONG-2020 EU Grant Agreement 824093 projects.

%\appendices

%\section{Appendices} \label{App:WhatGoes}
\bibliographystyle{JHEP}
\bibliography{PADMEMC}

\end{document}